\input colordvi
\documentclass[11pt]{article}
\usepackage{amsfonts,amssymb,a4}

\newcommand{\qed}{\hfill\rule{3mm}{3mm}}
\newtheorem{teorema}{Theorem}

\makeatletter
\@addtoreset{equation}{section}
\makeatother
\begin{document}


\voffset=-1.5truecm\hsize=16.5truecm    \vsize=24.truecm
\baselineskip=14pt plus0.1pt minus0.1pt \parindent=12pt
\lineskip=4pt\lineskiplimit=0.1pt      \parskip=0.1pt plus1pt

\def\ds{\displaystyle}\def\st{\scriptstyle}\def\sst{\scriptscriptstyle}


\let\a=\alpha \let\b=\beta \let\ch=\chi \let\d=\delta \let\e=\varepsilon
\let\f=\varphi \let\g=\gamma \let\h=\eta    \let\k=\kappa \let\l=\lambda
\let\m=\mu \let\n=\nu \let\o=\omega    \let\p=\pi \let\ph=\varphi
\let\r=\rho \let\s=\sigma \let\t=\tau \let\th=\vartheta
\let\y=\upsilon \let\x=\xi \let\z=\zeta
\let\D=\Delta \let\F=\Phi \let\G=\Gamma \let\L=\Lambda \let\Th=\Theta
\let\O=\Omega \let\P=\Pi \let\Ps=\Psi \let\Si=\Sigma \let\X=\Xi
\let\Y=\Upsilon


\global\newcount\numsec\global\newcount\numfor
\gdef\profonditastruttura{\dp\strutbox}
\def\senondefinito#1{\expandafter\ifx\csname#1\endcsname\relax}
\def\SIA #1,#2,#3 {\senondefinito{#1#2}
\expandafter\xdef\csname #1#2\endcsname{#3} \else \write16{???? il
simbolo #2 e' gia' stato definito !!!!} \fi}
\def\etichetta(#1){(\veroparagrafo.\veraformula)
\SIA e,#1,(\veroparagrafo.\veraformula)
 \global\advance\numfor by 1
 \write16{ EQ \equ(#1) ha simbolo #1 }}
\def\etichettaa(#1){(A\veroparagrafo.\veraformula)
 \SIA e,#1,(A\veroparagrafo.\veraformula)
 \global\advance\numfor by 1\write16{ EQ \equ(#1) ha simbolo #1 }}
\def\BOZZA{\def\alato(##1){
 {\vtop to \profonditastruttura{\baselineskip
 \profonditastruttura\vss
 \rlap{\kern-\hsize\kern-1.2truecm{$\scriptstyle##1$}}}}}}
\def\alato(#1){}
\def\veroparagrafo{\number\numsec}\def\veraformula{\number\numfor}
\def\Eq(#1){\eqno{\etichetta(#1)\alato(#1)}}
\def\eq(#1){\etichetta(#1)\alato(#1)}
\def\Eqa(#1){\eqno{\etichettaa(#1)\alato(#1)}}
\def\eqa(#1){\etichettaa(#1)\alato(#1)}
\def\equ(#1){\senondefinito{e#1}$\clubsuit$#1\else\csname e#1\endcsname\fi}
\let\EQ=\Eq


\def\\{\noindent}
\let\io=\infty

\def\VU{{\mathbb{V}}}
\def\ED{{\mathbb{E}}}
\def\GI{{\mathbb{G}}}
\def\Tt{{\mathbb{T}}}
\def\C{\mathbb{C}}
\def\LL{{\cal L}}
\def\RR{{\cal R}}
\def\SS{{\cal S}}
\def\NN{{\cal M}}
\def\MM{{\cal M}}
\def\HH{{\cal H}}
\def\GG{{\cal G}}
\def\PP{{\cal P}}
\def\AA{{\cal A}}
\def\BB{{\cal B}}
\def\FF{{\cal F}}
\def\TT{{\cal T}}
\def\v{\vskip.1cm}
\def\vv{\vskip.2cm}
\def\gt{{\tilde\g}}
\def\E{{\mathcal E} }
\def\I{{\rm I}}
\def\0{\emptyset}

\def\tende#1{\vtop{\ialign{##\crcr\rightarrowfill\crcr
              \noalign{\kern-1pt\nointerlineskip}
              \hskip3.pt${\scriptstyle #1}$\hskip3.pt\crcr}}}
\def\otto{{\kern-1.truept\leftarrow\kern-5.truept\to\kern-1.truept}}
\def\arm{{}}
\font\bigfnt=cmbx10 scaled\magstep1

\newcommand{\card}[1]{\left|#1\right|}
\newcommand{\und}[1]{\underline{#1}}
\def\1{\rlap{\mbox{\small\rm 1}}\kern.15em 1}
\def\ind#1{\1_{\{#1\}}}
\def\bydef{:=}
\def\defby{=:}
\def\buildd#1#2{\mathrel{\mathop{\kern 0pt#1}\limits_{#2}}}
\def\card#1{\left|#1\right|}
\def\proof{\noindent{\bf Proof. }}
\def\qed{ \square}
\def\reff#1{(\ref{#1})}
\def\eee{{\rm e}}

\title{Continuos particles in the Canonical Ensemble as an abstract polymer gas}
\author{
Thiago Morais$^{1,2}$ and
Aldo Procacci$^1$\\
\\
\small{$^1$ Departamento de Matem\'atica UFMG}
\small{ 30161-970 - Belo Horizonte - MG
Brazil}
\\
\small{$^2$ Departamento de Matem\'atica UFOP}
\small{ 35400-000 - Ouro Preto - MG
Brazil}
}
\maketitle

\def\be{\begin{equation}}
\def\ee{\end{equation}}
\vskip.5cm

\begin{abstract}
We revisit  the expansion recently proposed by Pulvirenti and Tsagkarogiannis for a system of $N$ continuous particles in the canonical ensemble.  Under the sole assumption that the particles interact via
a tempered and stable pair potential and are subjected to the usual free boundary conditions,  we show the analyticity of the  Helmholtz  free energy at low densities
and, using the Penrose tree graph identity, we  establish a lower bound for the convergence radius which happens to be identical to the lower bound of the convergence radius of the virial series
in the grand canonical ensemble established by Lebowitz and Penrose in 1964. We also show that the (Helmholtz) free energy can be written as a series
in power of the density whose $k$ order  coefficient coincides, modulo  terms ${o(N)/ N}$, with the $k$-order virial coefficient divided by $k+1$, according to its expression
in terms of the $m$-order (with $m\le k+1$) simply connected cluster integrals  first given by Mayer  in 1942. We finally give
 an upper bound for the $k$-order virial coefficient which slightly improves, at high temperatures, the  bound  obtained  by Lebowitz and Penrose.
 \end{abstract}
\numsec=1\numfor=1
\section{Introduction}
The rigorous approach  to systems constituted by a large number of  continuous classical interacting particles
has been  a deeply investigated subject during the last decades. In particular, the study of the low density phase, where
the behavior of the system should be  near to that of  an ideal  gas, has yielded some of the most impressive results in mathematical physics.
It is remarkable to note  that most of these results
have been obtained between  1962 and 1968.

As far as a system (gas) of continuous particles is concerned,  the Mayer series of the pressure (the pressure in powers of the fugacity) and the virial series of the pressure
(the pressure in powers of the density)  were known since the years around 1940. In particular, J. E. Mayer first gave the explicit expressions of the $n$-order
Mayer series coefficient in term of a sum  over connected graphs between $n$ vertices of cluster integrals, and of  the $n$-order virial series coefficient in term a sum  over two-connected graphs between $n+1$ vertices  of irreducible cluster integrals (see. e.g. \cite{MM} and references therein).
However the question regarding the convergence
of  these series remained unanswered  during the following two decades,  since a ``direct" upper bound
of the type $({\rm Const.})^n$ on these $n$ order  coefficients, which would have guaranteed analyticity of these series,  was generally  considered
rather prohibitive, due to the fact that the number of connected (or even two-connected) graph between $n$ vertices is too large
(i.e. order $C^{n^2}$ with $C>1$).\newpage

The rigorous analysis  of the Mayer series and the virial series of the pressure of the system of particles started to produce results with the work by Groeneveld \cite{gro62} in 1962, who first gave
a bound of the type $({\rm Const.})^n$ for the  $n$-order Mayer coefficient, but under the assumption that
the pair potential is non-negative. One year  later, Penrose \cite{Pe1, Pe2} and independently  Ruelle \cite{Ru1,Ru2} proved that the Mayer series of a system of continuous particles interacting via a stable and tempered pair potential (see ahead for the definition) is an analytic function for small values of the fugacity, as well providing
a lower bound for the convergence radius.  One year later Lebowitz and Penrose \cite{LP} obtained a lower bound for the convergence radius of the virial series (the pressure in function of the density).
These results were  all obtained ``indirectly", (i.e. not by trying to bound directly the  expressions of the coefficients as sum over connected graphs exploiting some cancelations) via the so called   Kirkwood-Salszburg Equations (KSE),
iterative relations between correlations functions of the system, and their possible use  towards the control of the convergence of the Mayer series and virial series was glimpsed since the forties
(see, e.g. \cite{May47,Ki} and reference therein).

An alternative method to KSE  was  proposed in the same years by
Penrose \cite{pen67} who proved the convergence of the Mayer series of a system of particles interacting via a pair potential
with a repulsive hard-core at short distance (but possibly attractive at large distance). To obtain this result he  rewrote the sum over connected graphs of the $n$-order Mayer coefficient
in terms of trees, by grouping together some terms, obtaining in this way the first example, as far as we know, of {\it tree graph identity} (TGI).
It was only a decade later  that  Brydges and Federbush \cite{BF} were able to provide, for the second time, a proof of the analyticity of the Mayer series for the pressure of a continuous gas by directly bounding its $n$-term coefficients via a new type of TGI.  Later,
this direct approach   based on TGI has been further developed and systematized by several authors
(see e.g.  \cite{bry84}, \cite{mal80}, \cite{BF}, \cite{AR}, \cite{PS}, \cite{pdls}, \cite{KKS}, \cite{uel04}).
Method based on KSE and TGI
are  part of the so called Cluster Expansion (CE) method.
TGI are nowadays  much more popular than the old KSE tools, mainly because of their flexibility and adaptability, but it is worths to remark that,
as far as continuous particle systems interacting via a stable and tempered pair potential are concerned, the bounds
given in \cite{Ru1,Ru2,Pe1,Pe2,LP} obtained via KSE have never been beaten.

An other very popular  tool in the framework of  CE methods, which we need to mention here, is the so called abstract polymer gas (APG).
The abstract polymer gas is basically a gas of subsets of some large set
called polymers which posses a fugacity and interact via a hard core (non overlapping) potential.
Such  model is in fact an extremely general  tool for investigating analyticity of
thermodynamic functions  of virtually any kind of lattice  system and its study has also a very long history which remounts to the sixties.
Indeed the Polymer gas, as a gas of subsets of the cubic lattice $\mathbb{Z}^d$, was originally  proposed in  the seminal papers  by Gallavotti and Miracle Sol\'e \cite{GM} (as a tool for the study
of the Ising model at low temperature) and Gruber and Kunz \cite{grukun71} (as a model in its own right),
where the analyticity of the pressure of such gas at low densities
is proved via KSE methods.  The same model was then studied also using TGI methods e.g.  by Seiler \cite{sei82} and
Cammarota \cite {cam82}. In 1986  Kotecky and Preiss \cite{kotpre86} proposed a downright ``abstract" polymer gas (polymers needed not to be subsets of an underlying set) and gave a proof of the convergence of the pressure
not based on the usual KSE or TGI cluster expansion methods. They also provided criterion
to estimate the convergence radius which improved those previously obtained via CE.
The proof of  the Koteck\'y-Preiss criterion was further simplified (and also slightly improved)
by Dobrushin \cite {dob96,dob96a} who reduced it to a simple inductive argument.  The beautiful inductive approach for the polymer gas formulated  by Dobrushin (which he called ``no cluster expansion approach") was
then generalized and popularized by  the work of Sokal \cite{sok01}  who also gave an extension of the APG convergence criterion for non-hard core repulsive pair interactions.
Finally, the robustness of
Cluster Expansion has been recently revalued in \cite{FP}
where the Dobrushin Kotecky-Preiss criterion has been improved via TGI arguments \cite{FP}. Moreover, always using TGI tools, further  generalizations of the APG convergence criterion  for non-hard core, non-repulsive pair interactions has been recently given in \cite{Pr1,Pr}, and \cite{PU}.

Turning back to continuous particle systems, as mentioned above,  all rigorous results about analyticity in the low density phase  were obtained in the Grand Canonical Ensemble.
Very recently  Pulvirenti and Tsagkarogiannis  \cite{PT} have obtained  a proof of analyticity of the free energy of a system of continuous particles in the Canonical Ensemble.
To get such a proof, they combined, within the cluster expansion methods, a standard Mayer expansion with the abstract polymer gas theory.
In this work  authors used the convergence  criterion for the abstract polymer gas given in \cite{bovzah00} and \cite{narolizah99}. However, this is not the best criterion in the literature.
In fact it is inferior not only to the recent Fernandez-Procacci criterion, but even  to the Kotecky-Preiss criterion. Moreover, for technical reasons, in \cite{PT}
the authors have used periodic  boundary conditions (instead of the usual free boundary conditions), and consequently they had to assume some further condition
on the pair potential beyond usual stability and temperness (see (2.3) in \cite{PT} and comments below).

In this note we revisit the calculations  in the canonical ensemble proposed by Pulvirenti and Tsagkarogiannis, but  under the sole assumption that particles interact via
a tempered and stable pair potential and are subjected to  the usual free boundary conditions. Using  the Fernandez-Procacci criterion to check the convergence of the polymer expansion,
we show the analyticity of the free energy at low densities in the canonical ensemble
and  establish a lower bound for the convergence radius which improves   the bound  given in \cite{PT} and  is identical to the lower bound of the convergence radius of the virial series
in the grand canonical ensemble established by Lebowitz and Penrose \cite{LP} in 1964.
We also show that the ( Helmholtz) free energy can be written as a series
in power of the density, whose $k$ order  coefficient coincides, up to  terms ${o(N)/N}$, with the $k$ order virial coefficient divided by $k+1$,
according to its expression in terms of (simply) connected cluster integrals  originally given by Mayer  in 1942  (formula (49)  in \cite{May42}). We finally give
 an upper bound for the $k$-order virial coefficient which slightly improves, at least at high temperatures, the  bound  obtained  by Lebowitz and Penrose in \cite{LP}.
\numsec=2\numfor=1
\section{Notations and results}
\def\xx{{\bf x}}
\def\pp{{\bf p}}
Throughout the paper, if $S$ is a set, then $|S|$ denotes its cardinality.  If $n$ is an integer then we will denote shortly $[n]=\{1,2,\dots,n\}$.

\subsection{Continuous particle system: Notations}
We  consider a system of  $N$ of  classical, identical
particles  enclosed in a  cubic box $\L\subset \mathbb{R}^d$ with
volume $V$ (and hence at fixed density $\r=N/V$). We suppose $N$ large (typically $N\approx 10^{23}$).
We denote by $x_i\in  \mathbb{R}^d$ the position vector  of the $i^{th}$
particle and by $|x_i|$ is modulus.
We assume that there are no particles outside $\L$ (free boundary conditions) and that these $N$
particles interact via a pair potential $V(x_i-x_j)$, so that the configurational energy
of the $N$ particles in the positions $(x_1,\dots ,x_N)\in \L^N$ is given by
$$
U(x_1,\dots, x_N)=\sum_{1\le i<j\le N} V(x_i-x_j)
$$
We make the following assumptions on the pair potential $V(x)$.
\begin{itemize}
\item[{A.}]Stability: there exists $B\ge 0$ such that,
for all $N\in \mathbb{N}$ and for all $(x_1,\dots, x_N)\in \mathbb{R}^{dN}$,
$$
\sum_{1\le i<j\le N} V(x_i-x_j) \ge -B N\Eq(2.6)
$$
\item[{B.}] Temperness:
$$
C(\b)=\int_{\mathbb{R}^d} |e^{-\b V(x)} -1| dx <+\infty \Eq(cbetat)
$$
\end{itemize}
The (configurational) partition  function  of this system in the canonical ensemble at fixed density $\r=N/V $ and fixed inverse temperature $\b\in \mathbb{R}^+$   is given by the following function
$$
Z_\L(\b,\r)={1 \over N!}\int_{\L}dx_1\dots\int_{\L} dx_N
 e ^{-\b \sum_{1\le i<j\le N}V(x_i-x_j) }
$$
The thermodynamics of the system can be derived from the partition function $Z_\L(\b,\r)$. In particular, the  Helmholtz free energy
per unit volume
of the system is  given by

$$
f(\b,\r)= \lim_{\L , N\to \infty\atop N/V =\r}f_\L(\b,\r)\Eq(aaa)
$$
where $\L\to \infty$ means that the size of the cubic box $\L$ goes to infinity and
$$
f_\L(\b,\r) = -{1\over \b V }\ln Z_\L(\b,\r) \Eq(fbL)
$$
We recall that the limit \equ(aaa) is known to exists if the pair potential $V(x)$ satisfies stability and temperness (see e.g. \cite{Ru}).

We will also make use in what follows of some notations concerning  the
system
in the Grand Canonical Ensemble. We first
recall the expression of the
Grand Canonical Partition function of the system enclosed in the volume $\L$, at fixed inverse temperature $\b$ and fixed fugacity $\l$.
$$
\Xi_\L(\b,\l)= \sum_{N\ge 0} {\l^N\over N!}\int_{\L}dx_1\dots\int_{\L} dx_N
 e ^{-\b \sum_{1\le i<j\le N}V(x_i-x_j) }
$$
with the $N=0$ term being equal to 1. The finite volume pressure of the system $P_\L(\b,\l)$ in the Grand Canonical ensemble is given by (see e.g. \cite{LP} or \cite{Ru})
$$
\b P_\L(\b,\l)={1\over V }\log \Xi_\L(\b,\l)= \sum_{n\ge 1} b_n(\b,\L){\l^n} \Eq(mayers)
$$
where $b_1(\b,\L)=1$ and, for $n\ge 2$,
$$
b_n(\b,\L)= {1\over V }{1\over n!}\int_{\L}d\xx_1
\dots \int_{\L} d\xx_n \sum\limits_{g\in G_{n}}
\prod\limits_{\{i,j\}\in E_g}\left[  e^{ -\b V(\xx_i -\xx_j)} -1\right]\Eq(ursm)
$$
where  $G_n$ is the set of all connected graphs with vertex set
$[n]$ and if $g\in G_n$ then its edge set is denoted by $E_g$. The r.h.s. of \equ(mayers) is known as Mayer series and the   term  $b_n(\b,\L)$  is the  $n$-order Mayer coefficient (a.k.a. $n$-order  connected cluster integral). We will need  in the next sections an upper bound for $|b_n(\b,\L)|$.
This term has been bounded several times in the literature
using both KSE or TGI methods. However,  as far as we know, the bound  obtained by
Penrose nearly fifty years ago  (see formula (6.12) in \cite{Pe1}) has never been beaten. The bound for $|b_n(\b,\L)|$ in \cite{Pe1} is, uniformly in $\L$ and  for all $n\ge 2$,
 as follows
$$
|b_n(\b,\L)|\le e^{2\b B (n-2)}n^{n-2} {[C(\b)]^{n-1}\over n!}\Eq(bmaru)
$$
where
$$
C(\b)=\int_{ \mathbb{R}^d} |e^{-\b V(x)}-1|dx\Eq(cbeta)
$$

An easy computation (see again \cite{LP}, \cite{Ru}) shows that the finite volume density $\r=\r_\L(\b,\l)$ of the system in the Grand  Canonical Ensemble is given by
$$
\r=
\sum_{n\ge 1} n b_n(\b,\L){\l^n} \Eq(density)
$$
So that one can eliminate $\l$ in \equ(mayers) and \equ(density) to obtain the so-called Virial expansion of the Pressure, i.e. the pressure in power of the density $\r=\r_\L(\b,\l)$, in the Grand canonical Ensemble
$$
\b P_\L(\b,\l)= \r - \sum_{k\ge 1} {k\over k+1} \b_k(\b,\L)\r^{k+1}\Eq(virial)
$$
where, as  shown  more than fifty years ago by Mayer (see e.g. \cite{MM} and reference therein)
$$
\b_k(\b,\L) ={1\over V }
{1\over k!}\int_\L dx_1\cdots \int_\L dx_{k+1} \sum_{g\in G^*_{k+1}}\prod_{\{i,j\}\in
E_g}[e^{-\b V(x_i,-x_j)}-1]\Eq(frmlls)
$$
with $G^*_{k+1}$ being the set of two-connected graphs  with  vertex set $[k+1]$. The  term $\b_k(\b,\L) $ is also known in the literature as
the irreducible cluster integral of order $k$.

As remarked in the introduction, the Mayer series in the r. h. s. of \equ(mayers) has been proved  to converge absolutely, uniformly in $\L$
\cite{Pe1, Pe2, Ru1,Ru2},
for any complex $\l$ inside the disk
$$|\l| <{1\over e^{2B+1} C(\b)}\Eq(radm)$$
where $B$  is the stability constant defined in \equ(2.6) and $C(\b) $ is the function defined in \equ(cbeta). Moreover Lebowitz and Penrose  \cite{LP} showed that the virial series in the  r.h.s. of \equ(virial) converges
 for all complex   $\r=\r_\L(\b,\l)$, uniformly in $\L$, inside the disk
$$
|\r|< g(e^{2\b B}){1\over e^{2\b B}C(\b)}\Eq(LeboPenr20)
$$
with
$$
g(u)= \max_{0<w<1} {[(1+u)e^{-w} -1]w\over u}\Eq(gdiu)
$$
It is important to stress one again that
the r.h.s. of \equ(radm), obtained in 1963,  and the r.h.s. of \equ(LeboPenr20), obtained one year later, still remain, as far as we know, the best  lower bounds for the convergence radius of the Mayer series of the pressure
and the convergence radius  of the virial series of the pressure, respectively, of a system  of continuous particles.

\subsection{Results: a Theorem in the Canonical Ensemble}
To state the results obtained in this paper (resumed in Theorem 1 below), let us introduce the following notations. Following \cite{PT}, we put
$$
Z_\L(\b,\r)= {V  ^{N} \over N!}\tilde Z_\L(\b,\r)
$$
where
$$
\tilde Z_\L(\b,\r)=  \int_{\L}{dx_1\over V }\dots\int_{\L} {dx_N\over V }
 e ^{-\b \sum_{1\le i<j\le N}V(x_i-x_j) }\Eq(Ztld)
$$
Let us define
$$
Q_\L(\b,\r)={1\over V } \log \tilde Z_\L(\b,\r)\Eq(cula)
$$
Then  the finite-volume  Helmholtz free energy $f_\L(\b,\r)$ defined in \equ(fbL) can be written as

$$
f_\L(\b,\r) = -{1\over \b}\left[{1\over  V }\ln\left({V ^N\over N!}\right) +Q_\L(\b,\r)\right]  \Eq(fbL2)
$$
where ${-1\over \b V }\ln{V ^N\over N!}$ is the  Helmholtz free energy of an ideal gas,
and ${-1\over \b}Q_\L(\b,\r)$
is the part of the   Helmholtz free energy due to the presence of the interaction $V(x)$.

\begin{teorema} Let $Q_\L(\b,\r)$ be defined as in \equ(cula), then the following statements are true.
\begin{itemize}
\item[i)]
It holds that
$$
Q_\L(\b,\r)=\sum_{k\ge 1}{\mathfrak{C}_k(\b,\L)\over k+1}\r^{k+1} \Eq(lnzt)
$$
where, for any fixed $k$,
$$
\lim_{N\to \infty} \mathfrak{C}_k(\b,\L)=\sum_{n=1}^k{(-1)^{n-1}} {(k-1+n)!\over k!} \sum_{{ \{m_2,\dots,m_{k+1}\}\atop m_i\in \mathbb{N}\cup\{ 0\},~\sum_{i=2}^{k+1} m_i=n}\atop
\sum_{i=2}^{k+1}(i-1)m_i=k}
\prod_{i=2}^{k+1} {[{b_{i}(\b,\L){i}}]^{m_i}\over m_i!}\Eq(ckappa0)
$$
with  the $b_{i}(\b,\L)$'s are the (simply connected) cluster integrals defined in \equ(ursm).

\item[ii)] Let
$$
\r^*_\b = { \mathcal{F}(e^{2\b B}) \over e^{2\b B}C(\b)}\Eq(LeboPenr)
$$
where $C(\b)$ is the function defined in \equ(cbeta) and
$$
\mathcal{F}(u)= \max_{a>0} \frac
{\ln[1+u(1-e^{-a})]}{e^a[1+u(1-e^{-a})]}
\Eq(gdiu0)
$$
Then the series in the r.h.s. of \equ(lnzt) converges absolutely, uniformly in $\L$, in the (complex) disk $|\r|\le \r^*_\b$.
\item[iii)] As soon as the density $\r$ is such that  $\r\le \r^*_\b$,  the factors $\mathfrak{C}_k(\b,\L)$ in r.h.s. of \equ(lnzt) admit the bound, uniformly in $\L$
$$
|\mathfrak{C}_k(\b,\L)|\le  \left[{1\over k+1}+(e^{a^*_\b}-1) e^{a^*_\b k}\right]  e^{2\b B(k-1)} {(k+1)^{k}\over k!}[C(\b)]^{k}\Eq(teo2)
$$
with $a^*_\b$ being the unique value of $a\in (0,\infty)$ such that the function in the r.h.s. of \equ(gdiu0)
reaches its minimum value (i.e. reaches the value $\r_\b^*$).
\end{itemize}
\end{teorema}

\vv
\\{\bf Remark 1}.
The theorem above immediately implies that the infinite volume   free energy  $f(\b,\r)$  defined in \equ(aaa) is also analytic in $\r$ in the same  disk $|\r|\le \r_\b^*$.

\vv
\\{\bf Remark 2}. It is not difficult to check that $\mathcal{F}(e^{2B\b})$ is an increasing  function of $\b$ with $\mathcal{F}(1)\approx 0,1448$ and $\lim_{\b \to\infty} \mathcal{F}(e^{2B\b})= e^{-1}$. Moreover $a^*_\b$ is a decreasing function
of $\b$ with $a^*_{\b=0}=0,426...$ and $\lim_{\b\to\infty }a^*_\b=0$.
We can thus compare this result with the best  lower bound for the  convergence radius of the virial series in the grand canonical ensemble (the pressure as a function of $\r$). Such bound was obtained by Lebowitz and Penrose in 1964. They found that the virial series is analytic in the open disk $|\r|<\mathcal{R}$ where (see formula (3.9) of \cite{LP})
$$
\mathcal{R}= {g(e^{2\b B})\over e^{2\b B}C(\b)}\Eq(LeboPenr2)
$$
with
$$
g(u)= \max_{0<w<1} {[(1+u)e^{-w} -1]w\over u}\Eq(gdiu2)
$$
It has been for us  quite surprising  to realize  that  $g(e^{2\b B})= \mathcal{F}(e^{2\b B})$,
so that the lower bound  \equ(LeboPenr) of the convergence radius of the  Helmholtz free energy as a function of the density in the canonical ensemble
 and the lower bound  \equ(LeboPenr2) of the convergence radius of the virial series in the grand canonical
ensemble given by Lebowitz and Penrose in  \cite{LP} are  in fact identical.

Indeed, first note that the function $G(w)=  {[(1+u)e^{-w} -1]w\over u}$ inside the max in r.h.s. of \equ(gdiu2)
 is positive only in the interval $w\in (0,\ln(1+u))$ and $G(0)=G(\ln(1+u))=0$. So we can rewrite
$$
g(u)= \max_{0<w<\ln(1+u)} {[(1+u)e^{-w} -1]w\over u}\Eq(gdiu3)
$$
On the other hand, via the  change of variables $w=\ln[1+u(1-e^{-a})]$  so that  $e^a=u/(u+1-e^w)$,
the r.h.s. of \equ(gdiu0) can be written

$$
 \FF(u)=\max_{a>0}\frac
{\ln[1+u(1-e^{-a})]}{e^a[1+u(1-e^{-a})]})= \max_{0\le w< \ln(1+u)}{w\over e^{w}}{(u+1-e^{w})\over u}= g(u)\Eq(fdiu3)
$$
\vv
\\{\bf Remark 3}.  Formula  \equ(ckappa0) is also quite remarkable since it  immediately implies  that $\mathfrak{C}_k(\b,\L)$ coincides, modulo terms of order ${o(N)\over N}$,  with the  $k$ order virial coefficient $\b_k(\b,\L)$. Indeed, r.h.s. of \equ(ckappa0) is, as was first
  shown by Mayer in 1942 \cite{May42}, the representation of the two-connected cluster integral $\b_k(\b,\L)$, defined in \equ(frmlls),
  in terms of the simply connected cluster integrals $b_i(\b,\L)$ ($i=1,\dots,k+1$), defined in \equ(ursm), (see  \cite{May42} formula 49, see also  formula (29) p. 319 of \cite{PB}). Therefore we  have, accordingly with \cite{PT}, that
$$
\mathfrak{C}_k(\b,\L)= \Big[{1 +{o(N)\over N}}\Big]\b_k(\b,\L)\Eq(frmllo)
$$

\vv
\\{\bf Remark 4}. We finally compare our bound for the virial coefficients \equ(teo2) with that obtained by Lebowitz and Penrose.  Their bound, as stated in formula (3.13) of \cite{LP} is
$$
k \b_k(\b,\L)\le \left[{(e^{2\b B} +1)C(\b)\over 0.28952}\right]^k\Eq(Lebp)
$$
On the other hand bound \equ(teo2) behaves asymptotically, taking for $a^*_\b$ its largest (and hence worst) value $a^*_{\b=0}=0,426$,  as
$$
Const.  \left[ { e^{2\b B}C(\b)\over 0.24026}\right]^{k}
$$
So our bound is asymptotically better than \equ(Lebp) for $\b$ small (i.e. high temperatures) and worst for $\b$ large (i.e. low temperatures).

\numsec=3\numfor=1
\section{Proof of Theorem 1}

Following  \cite{PT}, we can easily rewrite $\tilde Z_\L(\b,\r)$ defined in \equ(Ztld) as a partition function of a hard core polymer gas.
Indeed, using the Mayer trick we can rewrite
 the factor $e ^{-\b \sum_{1\le i<j\le N}V(x_i-x_j)} $ as
$$
e ^{-\b \sum_{1\le i<j\le N}V(x_i-x_j)} =
 \prod_{1\le i <j\le N}[{e ^{-\b V(x_i-x_j)} -1+1]} =
$$
$$
~=~ \sum_{\{R_1 ,\dots ,R_s\}\in\pi_N}\x(R_1)\cdots\x(R_r)
$$
where $\pi_N=$ set of all partitions of $[N]\equiv\{1,2,\dots, N\}$, and
$$
\x(R)~=~\cases{ 1 &if $|R|~=~1$\cr\cr \sum\limits_{g\in G_R}\prod\limits_{\{i,j\}\in
E_g}[e^{-\b V(x_i-x_j)}-1] &if $|R|\geq 2$\cr}
$$
with $G_R$  being  the set of  connected  graphs with vertex set $R$ and, given $g\in G_R$, $E_g$ denotes the set of  edges of $g$. Now define, for any $R\subset [N]$ such that $|R|\ge 2$,
$$
\z_{|R|}= \int_\L\dots \int_\L \prod_{i\in R} {dx_i\over V }\x(R)
\Eq(rhotilcan)
$$
Note that  $\z_{|R|}$ depends only of the cardinality of the polymer $R$ (the variables $\{x_i\}_{i\in R}$ are mute variables).  Observe also that
$$
\z_{n}=  {b_{n}(\b,\L)n!\over V ^{n-1}} \Eq(relrro)
$$
where, for any $n\ge 2$,  $b_n(\b,\L)$
is  the $n$ order coefficient of the Mayer series of the pressure defined in \equ(ursm).
It is now easy to check that
$$
\tilde Z_\L(\b,\r) =\Xi_{[N]}
=\sum_{n\ge 0}\sum_{\{R_1,\dots,R_n\}:~R_i\subset [N]\atop |R_i|\ge 2,\;R_i\cap R_j=\0}
{\z_{|R_1|}}\dots {\z_{|R_n|}}\Eq(GKpcan)
$$
with the $n=0$ term giving the factor 1.
So the partition function of a continuous gas
in the canonical ensemble is equal to the hard core polymer gas partition function $\Xi_{[N]}$
of a polymer gas in which the polymers are non overlapping subsets of the set $[N]$ with cardinality greater than one, and a polymer $R$ has activity $\z_{|R|}$ given by \equ(rhotilcan).
Note that the activity  $\z_{|R|}$
can be negative since, by \equ(relrro), its signal is the same  of the Mayer coefficient $b_{|R|}(\b,\L)$.

It is well known (see e.g. \cite{cam82}, \cite{PS}) that the logarithm of the hard core polymer gas partition function $\Xi_{[N]}$ can be written as
$$
\log \Xi_{[N]}= \sum_{n=1}^{\infty}{1\over n!}
\sum_{(R_{1},\dots ,R_{n})\in[N]^n}
\phi^{T}(R_1 ,\dots , R_n)\,{\z_{|R_1|}}\dots {\z_{|R_n|}}\Eq(6can)
$$
with
$$
\phi^{T}(R_{1},\dots ,R_{n})=\cases{1&if $n=1$\cr\cr
\sum\limits_{g\in G_{n}\atop g\subset G(R_1,\dots,R_n)}(-1)^{|E_g|} &if $n\ge 2$
}
\Eq(7can)
$$
where $G(R_1,\dots,R_n)$ is the graph with vertex set $[n]$ and edge set
$E_{G(R_1,\dots,R_n)}=\{\{i,j\}\subset [n]: R_i\cap R_j\neq \0\}$,  so  $\sum_{g\in{G}_n}$ is the sum over all connected
graphs with vertex set $[n]$ and $U(R,R')$ which are also subgraphs of $G(R_1,\dots,R_n)$. In conclusion we can  write
the function $Q_\L(\b,\r)$ defined \equ(cula) as
$$
Q_\L(\b,\r)={1\over V } \sum_{n=1}^{\infty}{1\over n!}
\sum_{(R_{1},\dots ,R_{n})\in[N]^n}
\phi^{T}(R_1 ,\dots , R_n)\,{\z_{|R_1|}}\dots {\z_{|R_n|}}\Eq(cula2)
$$

\subsection{Convergence criterion: proof of Theorem 1, part {\it ii}}

Once identity \equ(cula2) has been established, or, in other words, once  recognized that $Q_\L(\b,\r)$ is the logarithm, divided by $V $ of the partition function of a subset polymer gas (according  with the terminology used in \cite{bfp}),
we are in the position to prove
part ii of Theorem 1. By the Fernandez-Procacci criterion \cite{FP},
we have that $\log \Xi_{[N]}$ defined in \equ(6can) (and hence $Q_\L(\b,\r)$) can be written as an absolutely convergent series for all complex activities  $\z_{|R|}$   as soon as
$$
\sup_{i\in [N]}\sum\limits_{R\subset [N]:\; i\in R\atop |R|\ge 2}
|\z_{|R|}| ~  e^{a|R|}\le e^a-1\Eq(gkinducan)
$$
A few line  proof of this statement can also be found in \cite{bfp} (see there Theorem 2.4).
Now, since the sum the l.h.s. of \equ(gkinducan) does not depend on $i\in [N]$ and using also the fact that the activity
$\z_{|R|}$ depends only on the cardinality of the polymer $R$,  we can rewrite the condition \equ(gkinducan) as
$$
\sum_{m= 2}^{N}{e^{am}} C^\r_m
  \le e^a-1\Eq(gkinduca2n)
$$
where
$$
C^\r_m= |\z_m| {N-1\choose m-1} \Eq(cros)
$$

An upper bound for the activity $\z_{|R|}$ of a polymer $R$  is now the key ingredient to implement the convergence criterion \equ(gkinducan). Such a bound, recalling \equ(relrro),
 follows immediately from  the Penrose upper bound on the $n$-order Mayer coefficient given in \equ(bmaru).
So we have
$$
|\z_m|\le V ^{-m+1} e^{2\b B(m-2)} m^{m-2}[C(\b)]^{m-1}\Eq(asarm)
$$
and, recalling that $\r={N\over V }$, we get that $C^\r_m$ admits the following estimate
$$
C^\r_m\le \r^{m-1} e^{2\b B(m-2)}{ m^{m-2}\over (m-1)!}[C(\b)]^{m-1}\Eq(boundCr)
$$
where we have bound ${N-1\choose m-1}N^{-m+1}\le 1/(m-1)!$.

Hence,  the convergence condition   \equ(gkinduca2n)  is true if
$$
\sum_{m= 2}^{N}[\r \,{e^{a} C(\b)} e^{2\b B}]^{m-1} { m^{m-2}\over (m-1)!}
  \le e^{2\b B}(1-e^{-a})\Eq(gkinducan3)
$$
i.e. if
$$
\sum_{n= 1}^{\infty} {n^{n-1}\over n!}\Bigg[{e^{a}\over \k}\Bigg]^{n-1}
  \le 1+e^{2\b B}(1-e^{-a})\Eq(type)
$$
where  $\k= 1/(\r e^{2\b B}C(\b))  $.
Let now
$$
K^*=\min_{a\ge 0} \,\inf\biggl\{\k: \sum_{n=1}^{\infty}
\frac{n^{n-1}}{n!} \Bigl[\frac{e^a}{\k}\Bigr]^{n-1}\le
1+ e^{2\b B}\bigr(1-\,e^{-a}\bigr)\biggr\}
$$
As shown in \cite{Bo}  (see also \cite{FP2} and \cite{JPS}) $K^*$ can be written explicitly as
$$
K^*=\min_{a>0}\,
\frac{e^a [1+e^{2\b B}(1-e^{-a})]}
{\ln[1+e^{2\b B}(1-e^{-a})]}
$$
So \equ(gkinduca2n), and hence \equ(gkinducan), hold  for all complex $\r$ as soon as
$$
|\r|\le \r^*_\b
$$
where
$$
\r^*_\b= \FF(e^{2\b B}){1\over e^{2\b B}C(\b)} \Eq(canbou)
$$
and
$$
\FF(u)=  \max_{a>0}\,
\frac
{\ln[1+u(1-e^{-a})]}{e^a [1+u(1-e^{-a})]}
\Eq(fdiu)
$$
$\Box$

\subsection{Free energy  in powers of the density: proof of Theorem 1 part {\it i}}
To prove formula \equ(lnzt), we reorganize  the expansion \equ(6can), i.e.
$$
Q_\L(\b,\r)= {1\over V }\sum_{n=1}^{\infty}{1\over n!}
\sum_{(R_{1},\dots ,R_{n})\in[N]^n}
\phi^{T}(R_1 ,\dots , R_n)\,{\z_{|R_1|}}\dots {\z_{|R_n|}}
$$
as a power series in the density $\r$.
We use first of all the Penrose identity \cite{pen67,sok01,FP,Pr,JPS}  which states (we use the notation of \cite{FP})
 that
 $$
 \phi^{T}(R_1 ,\dots , R_n)= (-1)^{n-1}\sum_{\t\in T_n}\1_{\t\in P_{G(R_1,\dots, R_n)}}
 $$
where $P_{G(R_1,\dots, R_n)}$ are the set of Penrose trees of the graph $G(R_1,\dots, R_n)$ with vertex set $[n]$ and rooted in a fixed vertex of $[n]$, e.g., with root in the vertex  1. Thus

$$
Q_\L(\b,\r) = {1\over V } \sum_{n=1}^{\infty}{(-1)^{n-1}\over n!} \sum_{\t\in T_n}
\sum_{(R_{1},\dots ,R_{n})\in[N]^n}
\1_{\t\in P_{G(R_1,\dots, R_n)}}\,{\z_{|R_1|}}\dots {\z_{|R_n|}}\Eq(lognv0)
$$
We put
$$
\z_{s}= \r^{s-1}  \m_s \Eq(relrro2)
$$
where, recalling \equ(relrro),
$$
\m_s =   {b_s(\b,\L)s!\over N^{s-1}}\Eq(mus)
$$
Then
$$
Q_\L(\b,\r)= {\r\over N}\sum_{n=1}^{\infty}{(-1)^{n-1}\over n!} \sum_{\t\in T_n}\sum_{s_1,\dots s_n\atop sì\ge 2}
\sum_{(R_{1},\dots ,R_{n})\in[N]^n\atop |R_i|= s_i}
\1_{\t\in P_{G(R_1,\dots, R_n)}}\,{\z_{|R_1|}}\dots {\z_{|R_n|}}=
$$
$$
 = {\r\over N}\sum_{n=1}^{\infty}{(-1)^{n-1}\over n!} \sum_{s_1,\dots s_n\atop s_i\ge 2} \,{\z_{s_1}}\dots {\z_{s_n}} \sum_{\t\in T_n}
\sum_{(R_{1},\dots ,R_{n})\in[N]^n\atop |R_i|= s_i}
\1_{\t\in P_{G(R_1,\dots, R_n)}}=
$$
$$
 = {\r\over N}\sum_{n=1}^{\infty}{(-1)^{n-1}\over n!} \sum_{s_1,\dots s_n\atop s_i\ge 2} \r^{s_1-1}\m_{s_1}\dots \r^{s_n-1}\m_{s_n} \, \sum_{\t\in T_n}
\sum_{(R_{1},\dots ,R_{n})\in[N]^n\atop |R_i|= s_i}
\1_{\t\in P_{G(R_1,\dots, R_n)}}=
$$
$$
 = {\r\over N}\sum_{n=1}^{\infty}{(-1)^{n-1}\over n!} \sum_{k_1,\dots k_n\atop k_i\ge 1} \r^{k_1}\m_{k_1+1}\dots \r^{k_n}\m_{k_n+1} \sum_{\t\in T_n}
\sum_{(R_{1},\dots ,R_{n})\in[N]^n\atop |R_i|= k_i+1}
\1_{\t\in P_{G(R_1,\dots, R_n)}}\,=
$$
$$
 = {\r\over N}\sum_{n=1}^{\infty}{(-1)^{n-1}\over n!} \sum_{k\ge n}\r^k\sum_{k_1,\dots k_n:\,k_i\ge1\atop k_1+\dots +k_n=k}{\m_{k_1+1}}\dots {\m_{k_n+1}} \sum_{\t\in T_n}
\sum_{(R_{1},\dots ,R_{n})\in[N]^n\atop |R_i|= k_i+1}
\1_{\t\in P_{G(R_1,\dots, R_n)}}\,=
$$
$$
 = {\r\over N}\sum_{n=1}^{\infty}{(-1)^{n-1}\over n!} \sum_{k\ge n}\r^k\sum_{s_1,\dots s_n:\,s_i\ge 2\atop s_1+\dots +s_n=k+n} {\m_{s_1}}\dots {\m_{s_n}}\sum_{\t\in T_n}
\sum_{(R_{1},\dots ,R_{n})\in[N]^n\atop |R_i|= s_i}
\1_{\t\in P_{G(R_1,\dots, R_n)}}\,=
$$
$$
 = {\r\over N}\sum_{k\ge 1}\r^k\sum_{n=1}^k {(-1)^{n-1}\over n!}\sum_{s_1,\dots s_n;\,s_i\ge 2\atop s_1+\dots +s_n=k+n} {\m_{s_1}}\dots {\m_{s_n}}\sum_{\t\in T_n}
\sum_{(R_{1},\dots ,R_{n})\in[N]^n\atop |R_i|= s_i}
\1_{\t\in P_{G(R_1,\dots, R_n)}}\,=
$$
$$
 =\sum_{k\ge 1}{\r^{k+1}\over k+1}\sum_{n=1}^k {(-1)^{n-1}\over n!}\sum_{s_1,\dots s_n:\,s_i\ge 2\atop s_1+\dots +s_n=k+n} \prod_{i=1}^n
[{b_{s_i}(\b,\L){s_i}!}]{k+1\over N^{k+1}}\sum_{\t\in T_n}
\sum_{(R_{1},\dots ,R_{n})\in[N]^n\atop |R_i|= s_i}
\1_{\t\in P_{G(R_1,\dots, R_n)}}
$$
In conclusion we have proved formula \equ(lnzt) in Theorem 1, i.e. we have that
$$
Q_\L(\b,\r) = \sum_{k\ge 1}{\mathfrak{C}_k(\b,\L)\over k+1}\r^{k+1} \Eq(lnztOK)
$$
where
$$
\mathfrak{C}_k(\b,\L)=\sum_{n=1}^k{(-1)^{n-1}}W_n(k) \Eq(ckappa)
$$
and
$$
W_n(k)={k+1\over n!}\sum_{{s_1,\dots,s_n\atop s_i=2,3,\dots,}\atop s_1+\dots+s_n=k+n}\prod_{i=1}^n
[{b_{s_i}(\b,\L){s_i}!}] \PP(s_1,\dots,s_n)\Eq(wun)
$$
with
$$
\PP(s_1,\dots,s_n)={1\over N^{k+1}}
\sum_{\t\in T_n}
\sum_{(R_{1},\dots ,R_{n})\in[N]^n\atop |R_1|=s_1\dots ,|R_n|=s_n}
\1_{\t\in P_{G(R_1,\dots, R_n)}}\, ~\Eq(psss)
$$
We now have to show formula \equ(ckappa0).
We start  making
the following observation.
For fixed $k$ and $n$, the integers $s_1,\dots, s_n$ such that $s_1+\dots + s_n=k+n$ define uniquely  a $k$-tuple of integers
$\{m_2,\dots ,m_{k+1}\}$ such that $m_i\in\{0,1,2,\dots\}$ and $\{\# j\in[n]: s_j=i\}=m_i$ with the property that $\sum_{i=2}^{k+1}m_i=n$ and $\sum_{i=2}^{k+1}(i-1)m_i=k$ and so
$\prod_{j=1}^n[{b_{s_j}(\b,\L){s_j}!}]= \prod_{i=2}^{k+1} [{b_{i}(\b,\L){i}!}]^{m_i}$. Hence we can write

$$
W_n(k)= {k+1\over n!}\sum_{{\{m_i\}\equiv \{m_2,\dots,m_{k+1}\}\atop m_i\ge 0,~\sum_{i=2}^{k+1} m_i=n}\atop
\sum_{i=2}^{k+1}(i-1)m_i=k}
\prod_{i=2}^{k+1} [{b_{i}(\b,\L){i}!}]^{m_i} \sum_{{s_1,\dots,s_n\atop s_i\ge 2}\atop \{\# j\in[n]: s_j=i\}=m_i} \PP(s_1,\dots,s_n)\Eq(wnk77)
$$
Let us now calculate $\PP(s_1,\dots,s_n)$ under the conditions that   $s_1,\dots s_n$  is an  $n$-tuple of integers such that $s_i\ge 2$ and $s_1+\dots +s_n=k+n$  and such that it defines  the $k$-tuple of integers
$\{m_2,\dots ,m_{k+1}\}$. Recalling thus \equ(psss),
fix a tree $\t\in T_n$ and consider the factor
$$
w_\t=
\sum_{(R_{1},\dots ,R_{n})\in[N]^n\atop |R_1|=s_1\dots ,|R_n|=s_n}
\1_{\t\in P_{G(R_1,\dots, R_n)}}\, ~\Eq(vutau)
$$
This factor is clearly a polynomial in $N$ and, in view of the limit $N\to \infty$, we will retain only the
term of maximal degree in $N$. It is not difficult to see that once the tree $\t\in T_n$ has been fixed, the contribution
of the higher order in $N$ comes from the sum  over polymers $R_i$ submitted to the following prescriptions.
\begin{itemize}
\item
For any fixed vertex $i$ of the tree $\t$  different from the root with degree $d_i$
and $i_1,\dots,i_{d_i-1}$ children
sum over $R_{i_1},\dots R_{i_{d_i-1}}$ in such way that each one of the polymers $R_{i_1},\dots R_{i_{d_i-1}}$ shares exactly one vertex with $R_i$  and all these $d_i-1$ vertices are distinct
(so in particular we must have  $d_i-1\le |R_i|$, otherwise the contribution of this tree vanishes). This is because in such way the factors
$${N\choose |R_{i_1}|-1},\dots , {N\choose|R_{i_{d_i-1}}|-1}$$
 have the  maximal power   in $N$. For the root $1$ of $\t$,  with degree $d_1$ and $i_1,\dots,i_{d_1}$ children do analogously
(but this time it must hold $d_1\le |R_1|)$.
\item
In any fixed vertex $i$ of $\t$ we can choose  the remaining $|R_{i_j}|-1$ vertices in each polymer $R_{i_j}$ associated to the $i_j$ child of $i$ ($j=1,\dots, s_i$) among $N$ vertices; actually
the  $|R_{i_j}|-1$  vertices in each $R_{i_j}$ should be chosen  among $N_{i_j}<N$ vertices, where  $N-k-n\le N_{i_j}\le N-1$, according to the constraints imposed by the Penrose condition $\t\in P_{G(R_1,\dots, R_n)}$, but,
for $k$ fixed, we have that $k+n\le 2k$, so that $N(1-{2k\over N})\le N_{i_j}\le N(1-{1\over N})$.
\end{itemize}

These prescription  are exactly  the same conditions  used to calculate the r.h.s. of equation 3.20  in Lemma
3.4 of \cite{FP2}. Proceeding thus analogously to the computation explained in Lemma
3.4 of \cite{FP2} we have, for any $\t\in T_n$
$$
w_\t={N\choose s_1}{s_1\choose d_1} d_1!\prod_{i=2}^n {N\choose s_i-1}{s_i\choose d_i-1} (d_i-1)!  \Big[{1 +{o(N)\over N}}\Big] \1_{ d_1\le s_1,  d_i\le s_i+1}=
$$
$$
= {N^{s_1}\over s_1!}{s_1\choose d_1} d_1!\prod_{i=2}^n {N^{s_i-1}\over (s_i-1)!}{s_i\choose d_i-1} (d_i-1)!  \Big[{1 +{o(N)\over N}}\Big]\1_{ d_1\le s_1,  d_i\le s_i+1}=
$$
$$
= N^{k+1}{1\over s_1!}{s_1!\over  d_1!(s_1-d_1)!} d_1!\prod_{i=2}^n {1\over (s_i-1)!}{s_i!\over  (d_i-1)!(s_i-d_i+1)!} (d_i-1)!  \Big[{1 +{o(N)\over N}}\Big]\1_{ d_1\le s_1,  d_i\le s_i+1}=
$$
$$
=N^{k+1}{1\over (s_1-d_1)!}\prod_{i=2}^n{s_i\over (s_i-d_1+1)!}  \Big[{1 +{o(N)\over N}}\Big]\1_{ d_1\le s_1,  d_i\le s_i+1}
$$
where $d_i$ is the degree of vertices $i$ in $\t$ and we recall that for any tree $\t\in T_n$ we have that $d_1+\dots +d_n=2n-2$. So, summing over all trees $\t\in T_n$
and recalling Cayley formula,  we have that

$$
\PP(s_1,\dots,s_n)=\sum_{{d_1,\dots ,d_n:\; d_i\ge 1\atop d_1+\dots +d_n=2n-2}\atop 1\le d_1\le s_1,\; 1\le d_i\le s_i+1 } {(n-2)!\over \prod_{i=1}^n (d_i-1)!} {1\over (s_1-d_1)!}\prod_{i=2}^n{s_i\over (s_i-d_i+1)!}  \Big[{1 +{o(N)\over N}}\Big]
$$
$$
= {(n-2)!\over \prod_{i=1}^n(s_i-1)!}\sum_{{d_1,\dots ,d_n:\; d_i\ge 1\atop d_1+\dots +d_n=2n-2}\atop 1\le d_1\le s_1,\; 1\le d_i\le s_i+1 }  {(s_1-1)!\over (d_1-1)! (s_1-d_1)!}\prod_{i=2}^n{s_i!\over (d_i-1)!(s_i-d_i+1)!}  \Big[{1 +{o(N)\over N}}\Big]
$$
$$
= {(n-2)!\over \prod_{i=1}^n(s_i-1)!}\sum_{{l_1,\dots  l_n:\; l_i\ge 0\atop l_1+\dots +l_n=n-2}\atop 0\le l_1\le s_1-1,\; 0\le l_i\le s_i}
{(s_1-1)!\over l_1! (s_1-l_1-1)!}\prod_{i=2}^n{s_i!\over l_i!(s_i-l_i)!}  \Big[{1 +{o(N)\over N}}\Big]
$$
$$
= {(n-2)!\over \prod_{i=1}^n(s_i-1)!}\sum_{{l_1,\dots  l_n:\; l_i\ge 0\atop l_1+\dots +l_n=n-2}\atop 0\le l_1\le s_1-1,\; 0\le l_i\le s_i}
{s_1-1\choose l_1}\prod_{i=2}^n{s_i\choose l_i}  \Big[{1 +{o(N)\over N}}\Big]
$$
In conclusion we have obtained
$$
\PP(s_1,\dots,s_n) ={(n-2)!\over \prod_{i=2}^{k+1}[(i-1)!]^{m_i}}\sum_{{l_1,\dots  l_n:\; l_i\ge 0\atop l_1+\dots +l_n=n-2}\atop 0\le l_1\le s_1-1,\; 0\le l_i\le s_i}
{s_1-1\choose l_1}\prod_{i=2}^n{s_i\choose l_i}  \Big[{1 +{o(N)\over N}}\Big]\Eq(ppmi)
$$
We now show the identity
$$
 {\sum_{{l_1,\dots  l_n:\; l_i\ge 0\atop l_1+\dots +l_n=n-2}\atop 0\le l_1\le s_1-1,\; 0\le l_i\le s_i}
{s_1-1\choose l_1}\prod_{i=2}^n{s_i\choose l_i}} = {k-1+n\choose n-2}\Eq(combi0)
$$
Indeed, put $t_1=s_1-1$ and $t_i=s_i$ for all $i=2,\dots,n$.  So we need to prove that for any $n$-tuple  $t_1,\dots,t_n$ such that $\sum_{i=1}^n t_i=n+k-1$, $t_1\ge 1$ and $t_i\ge 2$ for
$i\ge 2$.
$$
  {\sum_{{l_1,\dots  l_n:\; l_i\ge 0\atop l_1+\dots +l_n=n-2}\atop 0\le l_i\le t_i}
\prod_{i=1}^n{t_i\choose l_i}}= {k-1+n\choose n-2} \Eq(combi)
$$
To prove this identity, suppose to have a set $V$ with $n+k-1$ objects and let $V=V_1\cup V_2\dots\cup V_n$
with $V_1,\dots, V_n$  being $n$ disjoint sets with cardinality $|V_1|=t_1,\dots, |V_n|=t_n$. Let $K_{n,k}$ be the number of ways
to pick up $n-2$ objects from the $n+k-1$ objects of $V$. Of course $K_{n,k}={k-1+n\choose n-2}$ and this is the r.h.s. of \equ(combi). On the other hand, since the sets $V_1,\dots, V_n$ form a partition of the set $V$
 we can also compute $K_{n,k}$  by choosing, for each $i\in [n]$,
 $l_i$ objects
from $V_i$ which is done in  ${t_i\choose l_i}$ ways, and then summing over all possible $n$-tuple $l_1,\dots,l_n$ under the constraint that $l_1+\dots+l_n=n-2$, getting in such way that
$K_{n,k}$ is also equal to the l.h.s. of \equ(combi). This prove \equ(combi) and hence \equ(combi0).
Putting this into \equ(ppmi) we have
$$
\PP(s_1,\dots,s_n) ={(n-2)!\over \prod_{i=2}^{k+1}[(i-1)!]^{m_i}} {k-1+n\choose n-2}(1+{o(N)\over N})\Eq(ppmi2)
$$
Note that \equ(ppmi2) implies that the factor $\PP(s_1,\dots,s_n)$ is, at least modulo terms of order ${o(N)\over N}$, a symmetric  function of $s_1,\dots,s_n$, i.e.  it depends only on $n,k$ and  numbers $\{m_2,\dots,m_{k+1}\}$. Actually one can easily realize that $\PP(s_1,\dots,s_n)$ is globally symmetric. Indeed, from definition \equ(psss) it immediately follows that
$\PP(s_1,\dots,s_n)$ is invariant under permutation of $s_2,\dots,s_n$ if the root of the Penrose trees is chosen to be $1$. Moreover, by the construction of the Penrose identity,  the r.h.s. of \equ(psss)
do not depend on the choice of the root, so that in the end $\PP(s_1,\dots,s_n)$ is actually invariant under permutation of $s_1,\dots,s_n$.

\\We can now plug  \equ(ppmi2) into \equ(wnk77) and we obtain
$$
W_n(k)= {k+1\over n!}\Big[1+{o(N)\over N}\Big]\!\!\sum_{{\{m_i\}\equiv \{m_2,\dots,m_{k+1}\}\atop m_i\ge 0,~\sum_{i=2}^{k+1} m_i=n}\atop
\sum_{i=2}^{k+1}(i-1)m_i=k}
\prod_{i=2}^{k+1} [{b_{i}(\b,\L){i}!}]^{m_i}  {(n-2)!\over \prod_{i=2}^{k+1}[(i-1)!]^{m_i}} {k-1+n\choose n-2}\!\!\!
\sum_{{s_1,\dots,s_n\atop s_i\ge 2}\atop \{\# j\in[n]: s_j=i\}=m_i}\!\!\!\!\!\!\!\!\!\!1
$$
So,  since
$$
\sum_{{s_1,\dots,s_n\atop s_i\ge 2}\atop \{\# j\in[n]: s_j=i\}=m_i}\!\!\!\!\!\!\!\!\!\!1~=~ {n!\over  \prod_{i=2}^{k+1}[m_i!]}
$$
we obtain
$$
W_n(k)= {(k+1)}(n-2)!\Big[1+{o(N)\over N}\Big]\sum_{{\{m_i\}\equiv \{m_2,\dots,m_{k+1}\}\atop m_i\ge 0,~\sum_{i=2}^{k+1} m_i=n}\atop
\sum_{i=2}^{k+1}(i-1)m_i=k}
\prod_{i=2}^{k+1} {[{b_{i}(\b,\L){i}}]^{m_i}\over m_i!} ~{k-1+n\choose n-2}
$$


$$
= {(k-1+n)!\over k!} [1+{o(N)\over N}] \sum_{{\{m_i\}\equiv \{m_2,\dots,m_{k+1}\}\atop m_i\ge 0,~\sum_{i=2}^{k+1} m_i=n}\atop
\sum_{i=2}^{k+1}(i-1)m_i=k}
\prod_{i=2}^{k+1} {[{b_{i}(\b,\L){i}}]^{m_i}\over m_i!}~~~~~~~~~~
$$
and so
$$
\mathfrak{C}_k(\b,\L)=[1+{o(N)\over N}]\sum_{n=1}^k{(-1)^{n-1}} {(k-1+n)!\over k!} \sum_{{\{m_i\}\equiv \{m_2,\dots,m_{k+1}\}\atop m_i\ge 0,~\sum_{i=2}^{k+1} m_i=n}\atop
\sum_{i=2}^{k+1}(i-1)m_i=k}
\prod_{i=2}^{k+1} {[{b_{i}(\b,\L){i}}]^{m_i}\over m_i!}\Eq(ckappa2)
$$
which concludes the proof of part i) of Theorem 1. $\Box$

As previously remarked,  r.h.s. of \equ(ckappa2) is, up to terms of order ${o(N)\over N}$, exactly the expression given in   formula  (49) of \cite{May42} (see also (29) p. 319 of \cite{PB}). We can thus conclude that
$\mathfrak{C}_k(\b,\L)$ is, up to terms of order ${o(N)\over N}$, the very same $k$ order virial coefficient as it is defined in formula (13.25) pag. 287 of \cite{MM}.
 So $\mathfrak{C}_k$ can also be written  in terms of a sum over two-connected graphs between $k+1$ vertices (see e.g. (13.25) in \cite{MM}).
Namely,
$\mathfrak{C}_k(\b,\L)= [1 +{o(N)\over N}]\b_k(\b,\L)$ where $\b_k(\\b,\L)$ is the virial coefficient defined in \equ(frmlls).

\subsection{Bound for the free energy: proof of Theorem 1, part {\it iii}}
Let us define the positive term series

$$
|Q|_\L(\b,\r) ={1\over V }\sum_{n=1}^{\infty}{1\over n!} \sum_{\t\in T_n}
\sum_{(R_{1},\dots ,R_{n})\in[N]^n}
\1_{\t\in P_{G(R_1,\dots, R_n)}}\,{|\z_{|R_1|}|}\dots {|\z_{|R_n|}|}\Eq(lognv)
$$
Then, clearly
$$
|Q_\L(\b,\L)|\le |Q|_\L(\b,\r)
$$
Now
$$
|Q|_\L(\b,\r)= {\r\over N}\sum_{n=1}^{\infty}{1\over n!} \sum_{\t\in T_n}\sum_{s_1,\dots s_n\atop sì\ge 2}
\sum_{(R_{1},\dots ,R_{n})\in[N]^n\atop |R_i|= s_i}
\1_{\t\in P_{G(R_1,\dots, R_n)}}\,{|\z_{|R_1|}|}\dots {|\z_{|R_n|}|}=
$$

$$
 = {\r\over N}\sum_{n=1}^{\infty}{1\over n!} \sum_{k\ge n}\r^k\sum_{s_1,\dots s_n\atop s_1+\dots s_n=k+n} \sum_{\t\in T_n}
\sum_{(R_{1},\dots ,R_{n})\in[N]^n\atop |R_i|= s_i}
\1_{\t\in P_{G(R_1,\dots, R_n)}}\,{|\m_{s_1}|}\dots {|\m_{s_n}|}=
$$
$$
 = {\r\over N}\sum_{k\ge 1}\r^k\sum_{n=1}^k {1\over n!}\sum_{s_1,\dots s_n\atop s_1+\dots +s_n=k+n} \sum_{\t\in T_n}
\sum_{(R_{1},\dots ,R_{n})\in[N]^n\atop |R_i|= s_i}
\1_{\t\in P_{G(R_1,\dots, R_n)}}\,{|\m_{s_1}|}\dots {|\m_{s_n}|}=
$$
$$
 =\sum_{k\ge 1}{\r^{k+1}\over k+1}\sum_{n=1}^k {1\over n!}\sum_{s_1,\dots s_n\atop s_1+\dots +s_n=k+n} \prod_{i=1}^n
[|{b_{s_i}(\b,\L)|{s_i}!}]{k+1\over N^{k+1}}\sum_{\t\in T_n}
\sum_{(R_{1},\dots ,R_{n})\in[N]^n\atop |R_i|= s_i}
\1_{\t\in P_{G(R_1,\dots, R_n)}}
$$
In conclusion we get
$$
|Q|_\L(\b,\r) = \sum_{k\ge 1}{|\mathfrak{C}|_k(\b,\L)\over k+1}\r^{k+1} \Eq(lnzt3)
$$
where
$$
|\mathfrak{C}_k|(\b,\L)=\sum_{n=1}^k |W|_n(k) \Eq(ckappaa)
$$
with
$$
|W|_n(k)={k+1\over n!}\sum_{{s_1,\dots,s_n\atop s_i=2,3,\dots,}\atop s_1+\dots+s_n=k+n}\prod_{i=1}^n
[{|b_{s_i}(\b,\L)|{s_i}!}] \PP(s_1,\dots,s_n)\Eq(wun2)
$$
Of course we have that
$$
|\mathfrak{C}_k(\b,\L)|\le |\mathfrak{C}_k|(\b,\L)\Eq(ofcor)
$$
We now obtain an upper bound for $|Q|_\L(\b,\r)$ as soon as $\r\le \r^*_\b$.
We start by writing, recalling \equ(lognv)

$$
|Q|_\L(\b,\r) = {1\over V }\sum_{n=1}^{\infty}{1\over n!} \sum_{\t\in T_n}
u_\t\Eq(pila)
$$
where
$$
u_\t=\sum_{(R_{1},\dots ,R_{n})\in[N]^n}
\1_{\t\in P_{G(R_1,\dots, R_n)}}\,{|\z_{|R_1|}|}\dots {|\z_{|R_n|}|}
$$
We first observe that the $n=1$ term in the sum in the r.h.s. of \equ(pila)  can be written as
$$
{1\over V }{1\over 1!} \sum_{\t\in T_1}
u_\t= {1\over V }\sum_{R\subset [N]\atop |R|\ge 2}|\z_{|R|}|= {1\over V }\sum_{s\ge 2}|\z_{s}| \sum_{R\subset [N]\atop |R|=s}1= {1\over V }
\sum_{s\ge 2} {N\choose s} |\z_{s}|=
$$
$$=
 {1\over V } \sum_{s\ge 2} {N\over s}{N-1\choose s-1} |\z_{s}| =  \r \sum_{s\ge 2} {1\over s} C^\r_s \Eq(uno)
$$
Now
following  \cite{FP2}, for a fixed $n\ge 2$ and  $\t\in T_n$,  we can bound
$$
|u_\t| \le \sum_{(R_{1},\dots ,R_{n})\in[N]^n}
\1_{\t\in P^*_{G(R_1,\dots, R_n)}}\,{|\z_{|R_1|}|}\dots {|\z_{|R_n|}|}
$$
where $P^*_{G(R_1,\dots, R_n)}$ are the weakly Penrose trees which are subsets of $G(R_1,\dots, R_n)$.
Using once again lemma 3.4 in \cite{FP2} we have, for any $\t\in P^*_{G(R_1,\dots, R_n)}$ and any $n\ge 2$,

$$
|u_\t|\le
\sum_{R_1\subset [N]\atop |R_1|\ge d_1\wedge 2}|\z_{|R_1|}|
{|R_1|\choose d_1}d_1! \prod_{i=2}^{n}\left[\sup_{j\in [N]}\sum_{R_i\subset [N],\,j\in R_i\atop |R_i|\ge (d_i-1)\wedge 2}
{|R_i|\choose d_i-1}(d_i-1)! |\z_{|R_i|}|\right]
$$
$$
\doteq w(d_1,\dots,d_n)~~~~~~~~~~~~~~~~~~~~~~~~~~~~~~~~~~~~~~~~~~~~~~~~~~~~~~~~~~~~~~~~~~
$$
where $d_i$ is the degree of vertex $i$ of $\t$. Hence, by  Cayley formula and definition \equ(cros)
$$
{1\over n!}\sum_{\t\in T_n}|u_\t|\le {1\over n!}
\sum_{d_1,\dots,\,d_n\atop d_1+\dots+d_n=2n-2}{(n-2)!\over \prod_{i=1}^n (d_i-1)!}~w(d_1,\dots,d_n)=
$$
$$
= {1\over n(n-1)}\sum_{d_1,\dots,\,d_n\atop d_1+\dots+d_n=2n-2}\sum_{R_1\subset [N]\atop |R_1|\ge d_1\wedge 2}|\z_{R_1}|
{|R_1|\choose d_1}d_1 \prod_{i=2}^{n}\left[\sup_{j\in [N]}\sum_{R_i\subset [N],\,j\in R_i\atop |R_i|\ge (d_i-1)\wedge 2}
{|R_i|\choose d_i-1} |\z_{R_i}|\right]
$$
$$
= {N\over n(n-1)}\sum_{d_1,\dots,\,d_n\atop d_1+\dots+d_n=2n-2}\sum_{s_1\ge d_1\wedge 2}{C^\r_{s_1}\over s_1}
{s_1\choose d_1}d_1 \prod_{i=2}^{n}\left[\sum_{s_i\ge (d_i-1)\wedge 2}
{s_i\choose d_i-1}C^\r_{s_i}\right]
$$
we now use the trick first used in \cite{PS} (see there section 3), so that, multiplying and dividing by $\a^{n-1}$ (with $\a>0$), we get,  for any $n\ge 2$,

$$
{1\over n!}\sum_{\t\in T_n}|u_\t|\le
{N \a^{-n+1}\over n(n-1)}
\sum_{d_1,\dots,\,d_n\atop d_1+\dots+d_n=2n-2}\sum_{s_1\ge d_1\wedge 2}{C^\r_{s_1}\over s_1}
{s_1\choose d_1}d_1 \a^{d_1} \prod_{i=2}^{n}\left[\sum_{s_i\ge (d_i-1)\wedge 2}
{s_i\choose d_i-1}C^\r_{s_i}\a^{d_i-1}\right]
$$
$$
~~~~~\le
{N\over \a^{n-1}n(n-1)}
\sum_{s_1,\dots,\,s_n\atop s_i\ge 2}{C^\r_{s_1}\over s_1}C^\r_{s_2}\cdots C^\r_{s_n}\sum_{d_1=1}^{s_1}
{s_1\choose d_1}d_1 \a^{d_1} \prod_{i=2}^{n}\left[\sum_{d_i-1=0}^{s_i}
{s_i\choose d_i-1}\a^{d_i-1}\right]
$$
$$
=
{N\over \a^{n-1}n(n-1)}
\sum_{s_1,\dots,\,s_n\atop s_i\ge 2}{C^\r_{s_1}\over s_1}C^\r_{s_2}\cdots C^\r_{s_n}s_1\a(1+\a)^{s_1-1}\prod_{i=2}^{n}\left[1+\a\right]^{s_i}~~~~~~~~~~~~~~~~
$$
$$
=
{N\over n(n-1)}\a \sum_{s_1\ge 2} C^\r_{s_1}(1+\a)^{s_1-1}
\Bigg[{1\over \a}\sum_{s\ge 2}(1+\a)^{s}C^\r_{s}\Bigg]^{n-1}~~~~~~~~~~~~~~~~~~~~~
$$
\underline{}Now, choosing $\a=e^{a^*_\b}-1$, we get, by the convergence criterion \equ(gkinduca2n), that  for $\r\le \r^*_\b$
$$
{1\over \a}\sum_{s\ge 2}(1+\a)^{s}C^\r_{s}\le 1
$$
So we get, for $\r\le \r^*_\b$
$$
{1\over V }\sum_{n\ge 2} {1\over n!}\sum_{\t\in T_n}|u_\t|\le
{\r}(e^{a^*_\b}-1)\sum_{s_1\ge 2} C^\r_{s_1}e^{a^*_\b(s_1-1)} \Eq(due)
$$
and hence, \equ(due) together with \equ(uno) yield the bound
$$
|Q|_\L(\b,\r)\le \r \sum_{s\ge 2}({1\over s}+ (e^{a^*_\b}-1) e^{a^*_\b(s-1)})C^\r_{s}
$$
Therefore,  recalling bound \equ(boundCr) for $C^\r_{s}$, we get
$$
|Q|_\L(\b,\r) \le \sum_{k\ge 1} {\r^{k+1}\over k+1}\left[{1}+(k+1) (e^{a^*_\b}-1) e^{a^*_\b k}\right]  e^{2\b B(k-1)} {(k+1)^{k-1}\over k!}[C(\b)]^{k} \Eq(bok)
$$
By comparing \equ(bok) with \equ(lnzt3) we immediately get
$$
|\mathfrak{C}_k|(\b,\L)\le \left[{1\over k+1}+(e^{a^*_\b}-1) e^{a^*_\b k}\right]  e^{2\b B(k-1)} {(k+1)^{k}\over k!}[C(\b)]^{k}
$$
and, by \equ(ofcor), the bound \equ(teo2) follows. $\Box$

\section*{Acknowledgments}
Aldo Procacci has   been partially supported by the Brazilian  agencies
Conselho Nacional de Desenvolvimento Cient\'{\i}fico e Tecnol\'ogico
(CNPq)  and  Funda{\c{c}}\~ao de Amparo \`a  Pesquisa do estado de Minas Gerais (FAPEMIG - Programa de Pesquisador Mineiro).


\begin{thebibliography}{99}


\bibitem{AR} A. Adbesselam and V. Rivasseau (1995): {\it Tree forests and jungles:
a botanical garden for cluster expansions in Constructive physics}, Proceedings, Palaiseau, France 1994, Lecture notes in physics n. 446.


\bibitem{bovzah00} A. Bovier and M. Zahradník (2000): A simple
  inductive approach to the problem of convergence of cluster
  expansions of polymer models. {\sl J. Statist.
  Phys.} {\bf 100}, 765--78.

\bibitem{bfp} R. Bissacot, R. Fern\'andez and A. Procacci (2010): {\it On the Convergence of Cluster Expansions for Polymer
Gases}, J. Statist. Phys., {\bf 139}, 598--617.

\bibitem{Bo} C. Borgs (2006): {\it  Absence of zeros for the chromatic polynomial on bounded degree graphs}. Combin.
Probab. Comput. {\bf 15}, 63--74.

\bibitem{BF} D. Brydges and P. Federbush (1978): {\it A new form of the Mayer expansion in classical statistical mechanics }. J. Math Phys. {\bf 19}, 2064 (4 pages).

\bibitem{bry84} D. C. Brydges (1984): A short cluster in cluster
  expansions.  In {\sl Critical Phenomena, Random Systems, Gauge
    Theories}, Osterwalder, K.  and Stora, R. (eds.), Elsevier,
  129--83.


\bibitem{cam82} C. Cammarota (1982): {\it Decay of correlations for
  infinite range interactions in unbounded spin systems}. { Comm.
    Math. Phys.}, {\bf 85}, 517--28.

\bibitem{dob96} R. L. Dobrushin (1996): {\it Estimates of semiinvariants
 for the {I}sing model at low temperatures}.  {Topics in
   Statistics and Theoretical Physics}, Amer. Math. Soc. Transl.,
  {\bf 177}, 59--81.

\bibitem{dob96a} R. L. Dobrushin (1996): Perturbation methods of the
 theory of {G}ibbsian fields.  In {\sl Ecole d'Et{\'e} de
  Probabilit{\'e}s de Saint-Flour XXIV -- 1994}, Springer-Verlag
 (Lecture Notes in Mathematics {\bf 1648}), Berlin--Heidelberg--New
 York, 1--66.

  \bibitem{FP} R. Fern\'andez and A. Procacci (2007): {\it Cluster
expansion for abstract polymer models. New bounds from an old
approach}, Commu. Math. Phys., {\bf 274},
123--140.

\bibitem{FP2} R. Fern\'andez and A. Procacci (2007): {\it Regions Without Complex Zeros
for Chromatic Polynomials
on Graphs with Bounded Degree}, Combin. Prob.  Comp. {\bf 17}, 225--238.

\bibitem{GM}
G. Gallavotti; S. Miracle-Sol\'e (1968): {\it Correlation functions for lattice systems}, Commun.
Math Phys. {\bf 7}, 274-288.


\bibitem{grukun71}
C. Gruber and H. Kunz (1971): {\it General properties of polymer systems}.
{ Comm. Math. Phys.}, {\bf 22}, 133--61.

\bibitem{JPS} B. Jackson, A. Procacci and A. D. Sokal (2013): {\it Complex zero-free regions at large $|q|$ for multivariate Tutte polynomials (alias Potts-model partition functions) with general complex edge weights}, J. Combin. Theory, Series B, {\bf 103}, 21--45.

\bibitem{Ki} J. G. Kirkwood (1946): {\it The statistical mechanical theory of transport processes}, J. Chem. Phys., {\bf 14}, 180-201

\bibitem{kotpre86} R. Koteck\'y and D. Preiss (1986): Cluster
  expansion for abstract polymer models.  {\sl Comm. Math. Phys.},
  {\bf 103}, 491--498.


\bibitem{KKS} T. Kuna, Yu. G. Kondratiev, and J. L. Da Silva (1998): {\it Marked Gibbs Measures via Cluster Expansion},
Methods Funct. Anal. Topology {\bf 4}, 50--81.

\bibitem{gro62}
  J. Groeneveld (1962): {\it Two theorems on classical many-particle
  systems}. {Phys. Lett.}, {\bf 3}, 50--51.

\bibitem{LP}  J. L. Lebowitz and O. Penrose (1964): {\it Convergence of Virial Expansions}, J. Math. Phys. {\bf 7}, 841-847.

\bibitem{mal80} V. A. Malyshev (1980): {\it Cluster expansions in lattice
  models of statistical physics and quantum theory of fields}.  {
    Russian Mathematical Surveys}, {\bf 35}, 1--62.

\bibitem{May42} J. E. Mayer  (1942): {\it Contribution to Statistical Mechanics}, J. Chem. Phys., {\bf 10}, 629--643.

\bibitem{MM} J. E. Mayer and M. G. Mayer (1940): {\it Statistical Mechanics}, John Wiley \& Sons, Inc.
London: Chapman \& Hall, Limited.

\bibitem{May47} J. E. Mayer  (1947):  Integral equations between distribution functions of molecules, J. Chem. Phye., {\bf 15}, 187--201.


\bibitem{mir00} S. Miracle-Sol\'e (2000): {\it On the convergence of cluster
  expansions}. { Physica A}, {\bf 279}, 244--249.

\bibitem{narolizah99} F. R. Nardi, E. Olivieri and M. Zahradník
  (1999): {\it On the Ising model with strongly anisotropic external field},
  { J. Statist. Phys.} {\bf 97}, 87--144.

\bibitem{PB} R. K. Pathria and P. D. Beale (2011): {\it Statistical mechanics, Third edition}, Elsevier, Amsterdam.

\bibitem{pfi91} Ch.-E. Pfister (1991): {\it Large deviation and phase
  separation in the two-dimensional Ising model}, {Helv. Phys. Acta},
{\bf 64}, 953--1054.

\bibitem{Pe1} O. Penrose (1963): {\it Convergence of Fugacity Expansions for Fluids and Lattice Gases}, Journal of Mathematical Physics {\bf 4},  1312 (9 pages).

\bibitem{Pe2} O. Penrose (1963): {\it  The  Remainder  in  Mayer's  Fugacity  Series}, J. Math. Phys. {\bf 4}, 1488 (7 pages).

 \bibitem{pen67} O. Penrose (1967): {\it Convergence of fugacity
 expansions for classical systems}.  In {\it Statistical
 mechanics: foundations and applications}\/, A. Bak (ed.),
 Benjamin, New York.

 \bibitem{PU} S. Poghosyan and D. Ueltschi (2009): {\it Abstract cluster expansion with applications to statistical mechanical systems}, J. Math. Phys. {\bf 50}, no. 5, 053509, (17 pp).
 \bibitem{Pr1} A. Procacci (2007):   {\it Abstract Polymer Models
with General Pair Interactions}, J. Stat Phys., {\bf 129}, 171--188.
 \bibitem{Pr} A. Procacci (2009): {\it Erratum and Addendum:``Abstract Polymer Models
with General Pair Interactions"}, J. Stat. Phys.,  {\bf 135}, 779--786.

\bibitem{pdls} A. Procacci, B. N. B. de Lima and B. Scoppola (1998): {\it A Remark on High Temperature Polymer
Expansion for Lattice Systems with
Infinite Range Pair Interactions}, Lett. Math. Phys., {\bf 45}, 303--322.

\bibitem{PS} A. Procacci and B. Scoppola (1999): {\it Polymer gas approach to
  $N$-body lattice systems}.  {\sl J. Statist. Phys.} {\bf 96}, 49--68.

\bibitem{PT}  E. Pulvirenti and D. Tsagkarogiannis (2012): {\it Cluster Expansion in the Canonical Ensemble}, Comm. math Phys.  {\bf 316}, Issue 2, pp 289--306.


\bibitem{Ru}  D. Ruelle (1969): {\it Statistical mechanics: Rigorous
    results}\/. W. A. Benjamin, Inc., New York-Amsterdam.

\bibitem{Ru1} D. Ruelle (1963): {\it Correlation functions of classical gases}, Ann. Phys., {\bf 5}, 109--120.

\bibitem{Ru2} D. Ruelle (1963): {\it Cluster Property of the Correlation Functions
of Classical Gases}, Rev. Mod. Phys., {\bf 36}, 580--584.


  \bibitem{sok01} A. D. Sokal (2001): {\it Bounds on the complex zeros of
(di)chromatic polynomials and Potts-model
partition functions}, { Combin. Probab. Comput.} {\bf 10}, 41--77.

\bibitem{sei82} E. Seiler (1982): {\sl Gauge Theories as a Problem of
    Constructive Quantum Field Theory and Statistical Mechanics},
  Lecture Notes in Physics {\bf 159}, Springer-Verlag,
  Berlin--Heidelberg--New York.


\bibitem{uel04} D. Ueltschi (2004): {\it Cluster expansions and
  correlation functions}. {Mosc.  Math. J.} {\bf 4}, 511--522.


\end{thebibliography}
\end{document}